# Adatom generated spatially oscillating strain fields on isotropic and other substrates

Wolfgang Kappus

Alumnus of ITP, Philosophenweg 19, D-69120 Heidelberg

wolfgang.kappus@t-online.de

v0.1: 2020-10-08

## Abstract

An extended elastic eigenvector approach for adatom interactions is applied to elastically isotropic substrates. The oscillating interactions between adatom monomers or dimers are caused by strain fields in the substrate. Expansive or contractive strain fields are surrounding adatoms in form of quasi circular rings. The exact shape of the ring structure depends on the elastic constants of the substrate and especially on the surface Brillouin zone.

Interactions on isotropic substrates are depicted and compared with an anisotropic example. Overlays of 2-D plots can help to identify adatom structures caused by elastic effects and to prepare numerical calculations of their elastic energies. The consequences of different sorts of adatoms with different interaction sign are sketched, large short distance attraction could occur in such cases.

Limitations of the method are discussed and possible applications are formulated.

## 1. Introduction

Different adatom configurations on substrates like Pd(001), Cu(111) and Cu(001) have recently been modeled by an an elastic eigenvector method with adatom pair- and multisite interactions [1, 2, 3]. The following sections rely on the details given there.

The interaction of adatoms or dimers is mediated by strain fields caused by stress to their vicinity. Spatial oscillations are the consequence of the sharp cutoff of a wave vector integral [1].

Assuming expansive stress an adatom or dimer exerts to the substrate, repulsive or attractive interactions are a consequence of expansive or contracting strain fields in their neighborhood.

On isotropic substrates the eigenvector equation [1] has a closed solution and allows a more intuitive interpretation. This will allow also a better understanding of the surface Brillouin zone's influence. Tungsten will be used as substrate example, the differences on various surfaces will be shown. Cu(001) is used to show the effect of elastic anisotropy.

Interaction- resp. strain plots will be shown and the potential of templates to search elastic effects in adatom configurations will be discussed.

The possibility of short range attractive elastic interactions of different adatom types will be raised; potential effects in catalysis are sketched.

Open questions are formulated.

# 2. Elastic adatom interaction

From [3] the elastic interaction between two adatoms $V_{11}$ or between two dimers $V_{22}$

$$V_{kl}(s, \chi) = (2\pi)^{-1} \sum_p \omega_{kl,p} \cos(p\chi) \cos(p\pi/2)\, 2^{-1-p} \kappa_{BZ}{}^3 (s\,\kappa_{BZ})^p\, \Gamma\!\left(\frac{3+p}{2}\right) *$$

$$_1F_2\!\left(\left(\frac{3+p}{2}\right), \left(\frac{5+p}{2}, 1+p\right), -\frac{1}{4} s^2 \kappa_{BZ}{}^2\right) X(s), \qquad (2.1)$$

with the heuristic factor

$$X(s) = (s/s_0)^{-3/2} \qquad (2.2)$$

is recalled. The factor in Eq. (2.2) ensures proper mesoscale $s^{-3}$ decay; $s_0$ is the lattice constant of Cu.

In Eq.(2.1) $\Gamma(p)$ denotes the Gamma function and $_1F_2(a,b_1,b_2,s)$ the generalized hypergeometric function, $\kappa_{BZ}$ is the wave vector distance between the origin and the surface Brillouin zone in $\chi$ direction. $\omega_{kl,p}$ are eigenvalues for the 2 interaction types with $p$ resulting from a series expansion. $p$ takes the values 0, 4, 8 in the monomer case and 0, 2, 4 in the dimer case. The eigenvalues $\omega_{kl,p}$ are proportional to the product of stress parameters $P_k P_l$ and inversely proportional to the elastic constant $c_{44}$ defining a dimensionless constant $\hat{\omega}_{kl,p}$ by

$$\omega_{kl,p} = \hat{\omega}_{kl,p}\, P_k P_l / c_{44}\,. \qquad (2.3)$$

$P_1$ is the strength of isotropic stress an adatom exerts to the surface, $P_2$ is the strength of anisotropic stress a dimer exerts to the surface [1].

## 2.1. Elastic adatom interaction on isotropic substrates

Tables 1 shows the dimensionless coefficients $\hat{\omega}_{kl,p}$ for an isotropic substrate with $c_{11}-c_{12}-2c_{44}=0$.
The $k=l=1$ coefficients belong to a monomer-monomer (pair) interaction $V_{ij}$ in Eq. (2.1).
The $k=1$, $l=2$ coefficients belong to monomer-dimer (trio) interactions.
The $k=2$, $l=2$ coefficients belong to (quarto) interactions of two parallel dimers.
The $k=2$, $l=3$ coefficients belong to (quarto) interactions of two dimers spanning an angle of 90°.

| k | l | $\omega_{kl,0}$ | $\omega_{kl,2}$ | $\omega_{kl,4}$ |
|---|---|---|---|---|
| 1 | 1 | $-c_{11}/(c_{11}-c_{44})/2$ | □ | □ |
| 1 | 2 | $-\sqrt{2}\,c_{11}/(c_{11}-c_{44})/4$ | $-\sqrt{2}\,c_{11}/(c_{11}-c_{44})/4$ | □ |
| 2 | 2 | $(-5c_{11}+2c_{44})/(c_{11}-c_{44})/8$ | $-c_{11}/(c_{11}-c_{44})/2$ | $(c_{11}-2c_{44})/(c_{11}-c_{44})/8$ |
| 2 | 3 | $(c_{11}-2c_{44})/(c_{11}-c_{44})/8$ | 0 | $-(c_{11}-2c_{44})/(c_{11}-c_{44})/8$ |

Table 1. Coefficients $\omega_{kl,p}$ (in $P_k P_l/c_{44}$ units) for elastic isotropic substrates.

## 2.2. Surface Brillouin zone

Due to the oscillating interaction the choice of the surface Brillouin zone shape in Eq. (2.1) has a significant influence. Like in [3] a square approximation

$$\kappa_{BZ}(\chi) = \kappa_{BZ0}(15/16 + \cos(4\chi)/16) \ . \quad (2.4)$$

is selected, ensuring $\kappa_{BZ}(0) = \kappa_{BZ0} = fbz*2\pi/s_0$. This choice reflects the symmetry on the (001) surface, the value 16 is chosen to give zero curvature in the (100) direction and the factor *fbz*=0.97.

In case the substrate has no crystalline order the surface Brillouin zone will be assumed a circle with

$$\kappa_{BZ}(\chi) = 2\pi/s_0. \quad (2.5)$$

# 3. Surface strain fields

The elastic interactions of Eq. (2.1) are closely related to the strain field around a reference adatom.

If an adatom or dimer exerts expansive stress to an isotropic substrate, it creates alternating rings of expansive or contracting strain, starting with an expansive disk in its direct neighborhood. Attraction occurs when neighboring adatoms or dimers (also exerting expansive stress) are located within an expansive ring.

Repulsion occurs when neighboring adatoms or dimers (also exerting expansive stress) are located within an contracting ring.

In Fig. 1 various strain fields resp. various interactions from Eqs. (2.1) to (2.5) are shown. All plots show a negative hole at the distance $s \gtrsim 0$, high maxima near $s \approx 0.5 s_0$ ( both not relevant as adatom positions), a steep descent at $s \lesssim s_0$, a minimum near $s \approx 1.3 s_0$ followed by a maximum. Oscillations of interaction values near zero are strongly decreasing with increasing distance $s$. The left column (Fig. 1.a, c, e) shows pair interactions $V_{11}$, the right column (Fig. 1.b, d, f) shows interactions $V_{22}$ of y-directed parallel dimers. The 3 rows indicate different substrates; the first row (Fig. 1.a, b) shows tungsten with an isotropic circular surface Brillouin zone according to Eq. (2.5); the second row (Fig. 1.c, d) shows W(001), an elastically isotropic substrate; the third row (Fig. 1.e, f) shows Cu(001), an elastically anisotropic substrate; both surface Brillouin zones according to Eq. (2.4). The interactions in the first row are similar to those on W(111).

If the reference adatom or reference dimer exerts expansive stress to the surface then the first minimum region shows expansive strain.

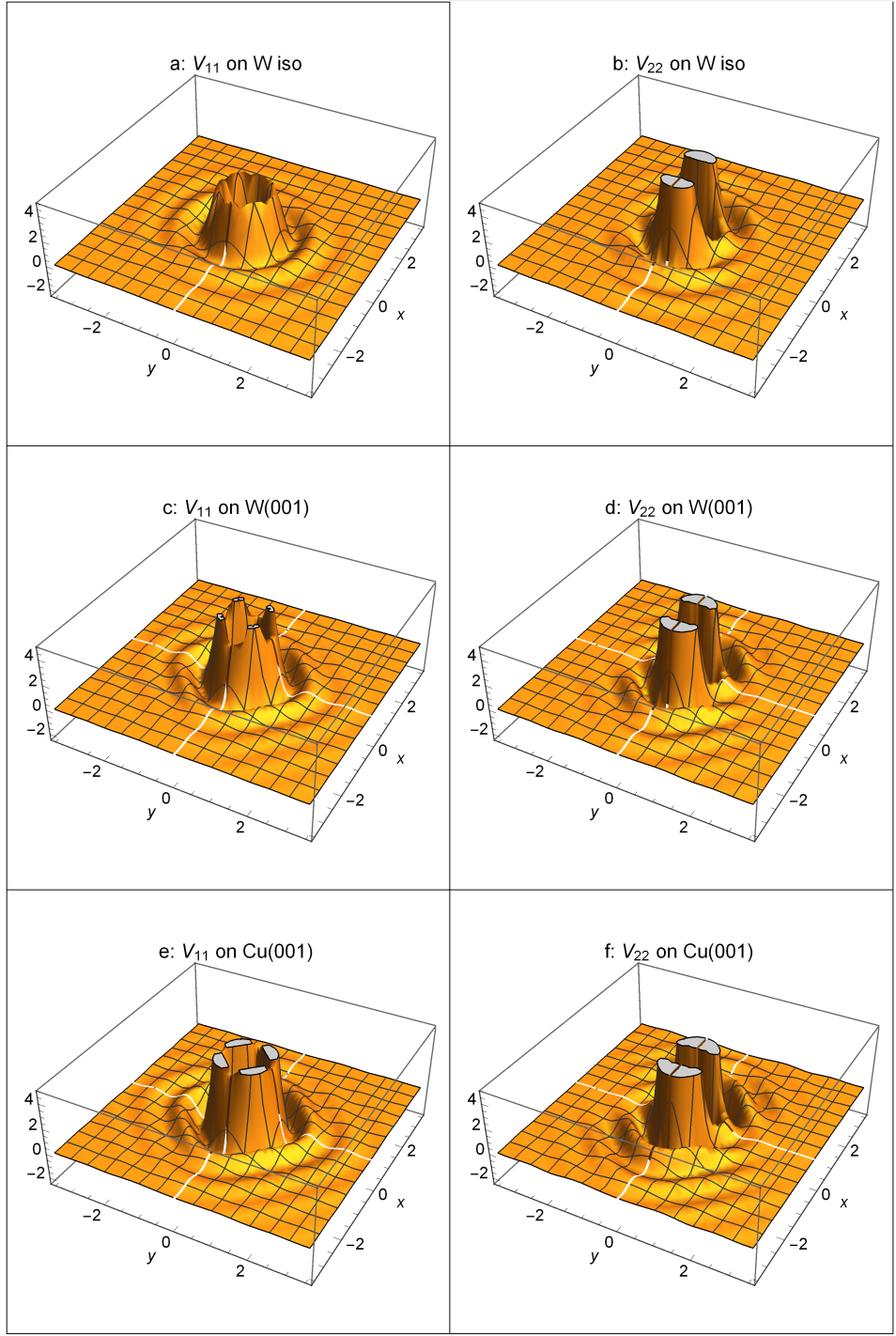

Fig. 1. Pair interaction $V_{11}$ (in $P_1^2$ units) and dimer-dimer interaction $V_{22}$ (in $P_2^2$ units) on a, b: W, an isotropic substrate, with isotropic Brillouin zone c, d: W(001) with anisotropic Brillouin zone e, f: Cu(001) with anisotropic Brillouin zone.

Because the 3D plots in Fig. 1 do not allow to extract details, interaction values at the first minimum and maximum in 3 directions are shown in Tab. 2. There are no fundamental differences among $V_{11}$ values and $V_{22}$ values for the 3 different surfaces.

| ☐ | Figure 1. | a | b | c | d | e | f |
|---|---|---|---|---|---|---|---|
| direction | location | $V_{11}$ W iso | $V_{22}$ W iso | $V_{11}$ W (001) | $V_{22}$ W (001) | $V_{11}$ Cu (001) | $V_{22}$ Cu (001) |
| x | min | −0.6 | −0.15 | −0.6 | −0.15 | −1.2 | −0.2 |
| y | min | −0.6 | −1.1 | −0.6 | −1.1 | −1.2 | −1.9 |
| diag | min | −0.6 | −0.75 | −1.0 | −1.2 | −1.1 | −1.4 |
| x | max | 0.25 | 0.05 | 0.2 | 0.05 | 0.4 | 0.1 |
| y | max | 0.25 | 0.5 | 0.2 | 0.55 | 0.4 | 0.85 |
| diag | max | 0.25 | 0.3 | 0.4 | 0.45 | 0.6 | 0.75 |

Table 2. $V_{11}$ and $V_{22}$ interaction values on different surfaces taken from Fig. 1 at the first minimum and the first maximum in x-, y-, diagonal direction.

# 4. Discussion

Details of adatom interactions can lead to quite different adatom structures. The minimum total energy of an adatom configuration will in equilibrium determine the structure. Experimentalists may look for a simple method to identify the potential occurrence of elastic effects; the findings of section 3 will be used to propose a template scheme for this purpose.

### 4.1. Template scheme

Adatom areas, i.e. 2-D versions of Fig. 1, can be put like puzzle pieces on a plane grid representing adatom positions. The centers of those pieces must be located on the grid intersections. An attractive elastic interaction is given if a center of such a piece is located within a minimum area of a neighboring piece. A realistic adatom structure would result if most of the centers of representatives are located within minimum areas.

With the help of computer aided design (CAD) systems the search for realistic structures could be simplified.

### 4.2. Opposite sign of stress parameters

If two different sorts of adatoms or dimers are present on a surface, they exert different stress on their neighborhood. If the sign of their stress parameters differs, i.e. if one exerts contractive stress and the other exerts expansive stress, then according to Eqs. (2.3) and (2.1) the sign of their interaction is changed. Consequently Fig. 1 is mirrored on the $z$=0 plane.

If the size of such unequal adatoms is not larger than the lattice constant $s_0$ of the substrate, then the large repulsion of equal adatoms in the $s<s_0$ regions of Fig. 1 is converted into large attraction. This could be an explanation for catalytic activity.

### 4.3. Open questions

The current base for the elastic eigenvector model lacks a microscopic foundation [4]. Three applications are just a start, more adatom systems with elastic features would solidify the approach. The present continuous elastic model is valid for the mesoscale; it cannot be expected to cover very small distances.

The current model heavily relies on the surface Brillouin zone shape and the $\kappa$ integration cutoff. The shape is not the natural choice and spatial oscillations are a consequence of the sharp cutoff, the exact form of which needs further investigation.

Elastic multisite effects, caused by adatom dimers, could occur frequently but may be hidden by dominating pair interactions.

# 5. Summary

The elastic eigenvector method has a closed solution for the elastic energy of adatoms on isotropic substrates like tungsten. The energy's spatial oscillations are shown in 3-D plots and related to expansive and contractive strain areas. A template scheme to detect elastic effects is proposed. The consequences of different sorts of adatoms for the interaction sign are sketched.